\documentclass{aa}
\usepackage{amsmath,amssymb}
\usepackage{graphics}

\begin{document}
\thesaurus{03(12.07.1; 11.04.1; 09.04.1; 03.13.4)}
\title{Redshift estimate of a gravitational lens from the observed
reddening of a multiply imaged quasar}
\author{C. Jean \and J. Surdej\thanks{Directeur de Recherches du FNRS (Belgium)}}
\offprints{C. Jean, jean@astro.ulg.ac.be}
\institute{Institut d'Astrophysique, Universit\'e de Li\`ege,\\ Avenue de Cointe 5, B-4000 Li\`ege, Belgium}
\date{Received / Accepted }
\titlerunning{Redshift estimate of a lens}
\authorrunning{C. Jean \& J. Surdej}
\maketitle

\begin{abstract}

Light rays from a multiply imaged quasar usually sample different path lengths
across the deflector. Extinction in the lensing galaxy may thus lead
to a differential obscuration and reddening between the observed macro-lensed
QSO images. These effects naturally depend on the precise shape of the
extinction law and on the redshift of the lens. By means of numerical
Monte-Carlo simulations, using a least-squares fitting method and assuming an
extinction law similar to that observed in the Galaxy, we show how accurate
photometric observations of multiply imaged quasars obtained in several
spectral bands could lead to the estimate of the lens redshift, irrespective
of the visibility of the deflector. Observational requirements necessary to
apply this method to real cases are thoroughly discussed. If extinction laws
turn out to be too different from galaxy to galaxy, we find out that more
promising observations should consist in getting high signal-to-noise low
resolution spectra of at least three distinct images of a lensed quasar,
over a spectral range as wide as possible, from which it should be
straightforward to extract the precise shape of the redshifted extinction law.
Very high signal-to-noise, low spectral resolution, VLT observations of
\object{H1413+117} and \object{MG~0414+0534} should enable one to derive such
a redshifted extinction law.


\end{abstract}
\section{Introduction}
For the case of several known gravitational lens systems, the deflector is
either not yet detected or it is so faint that its redshift remains unknown
(see the Gravitational Lensing web page at \texttt{http://vela.astro.ulg.ac.be/grav\_lens/}).
Measurement of the deflector redshift is mandatory in order to properly model
the corresponding gravitational lens systems. Usually, the lens redshift is
determined from spectroscopic observations of either emission lines due to the
lensing galaxy or of galaxy absorption lines detected in the spectra of the QSO
images. In this work, we present an alternate method to derive the redshift of 
a too faint and/or an invisible lens. First, we describe the method which is 
based upon the fitting of photometric observations with a redshifted 
extinction model and we then show results from Monte Carlo simulations. The 
latter turn out to be very useful in order to assess the errors affecting 
the estimate of the lens redshift. Next, we attempt to apply this method 
to several known gravitational lens systems and outline all observational 
requirements needed in order to derive successful results. Finally, we propose 
a generalization of this method. Conclusions form the last section.
\section{Description of the method}
\subsection{Generalities}
\label{generalities}
\begin{figure}[h]
\resizebox{\hsize}{!}{\includegraphics{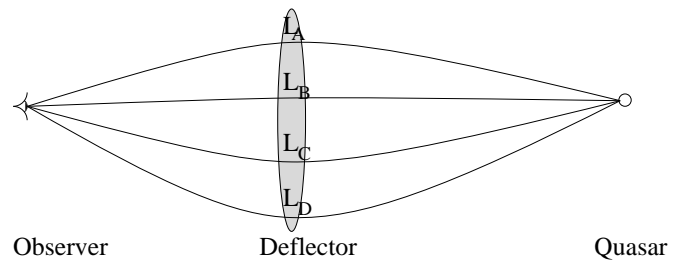}}
\caption{Schematic representation of a multiply imaged quasar.
$L_A-L_D$ denote different path lengths across the deflector.}
\label{schema}
\end{figure}

Figure \ref{schema} depicts a gravitational lens system. The light rays
emitted from the quasar are deflected by the lensing galaxy and the resulting
macro images are reddened because of extinction effects in the deflector. Due
to the different path lengths across the galaxy, differential reddening
results between the macro lensed components.

The flux received by the observer from component $i$ at a wavelength
$\lambda$ is:
\begin{equation}
F_{\lambda}^i = F_{\lambda}^0 A_i e^{-\text{Ext}(\lambda',R_V)L_i}
\label{flux}
\end{equation}
where $F_{\lambda}^0$ is the flux that would be observed from the quasar in
the absence of lensing, $A_i$ the macro lens amplification of component $i$,
Ext($\lambda',R_V$) the Galaxy dust opacity at the wavelength
$\lambda'=\frac{\lambda}{1+z_l}$ ($z_l$ being the lens redshift) and
$R_V = \frac{A(V)}{E(B-V)}$ (see Cardelli \textit{et al.} \cite{Cardelli}), and $L_i$ the
path length of component $i$ through the deflector.

When deriving Eq. (\ref{flux}), we have of course implicitly assumed that the
macro lensed images are not contaminated by light from the lensing galaxy,
that they are not affected by microlensing effects, that no intrinsic colour
variations exist between the lensed components, that the law of extinction 
(first assumed to be alike that of the Galaxy) is
similar all through the deflector and that the observed reddening takes place
in a single lensing plane. A check on the applicability of most of these
conditions can be made as suggested in section \ref{colour-colour_diagrams}.

The colour of component $i$, between the wavelengths $\lambda_1$ and
$\lambda_2$, is then merely given by:
\begin{equation}
m_{\lambda_1,i}-m_{\lambda_2,i} = -2.5\log\left(\frac{F_{\lambda_1}^0 \, e^{-\text{Ext}(\lambda_1',R_V)L_i}}
{F_{\lambda_2}^0 \, e^{-\text{Ext}(\lambda_2',R_V)L_i}}\right).
\end{equation}
Having chosen one of the macro-lensed images as a reference component (i.e. $i=\text{ref}$),
the relative colour of component $j$ (i.e. $i=j$) may be expressed as:
\begin{multline}
(m_{\lambda_1,j}-m_{\lambda_2,j})-(m_{\lambda_1,\text{ref}}-m_{\lambda_2,\text{ref}}) =\\
\frac{2.5}{\ln(10)} \, \mathcal{L}_j \, [\text{Ext}(\lambda_1',R_V) - \text{Ext}(\lambda_2',R_V)]
\label{rel_colour}
\end{multline}
where $\mathcal{L}_j = L_j - L_{\text{ref}}$ represents the relative path
length of component $j$.
\subsection{Observational constraints from the measurement of colours}
\label{constraints}
If we assume that the gravitational lens system consists of $N$ macro lensed
components and that observations are being carried out through $M$ filters, we
easily find that there are $M-1$ independent colours for each of the $N-1$
components (the reference component is of course not included). Thus, there
results a set of $(M-1)(N-1)$ equations alike Eq. (\ref{rel_colour}).

Besides, there are 2 unknowns which do not depend on the numbers of filters and
components: $z_l$ and $R_V$. Moreover, for each of the $N-1$ components, there
is one additional unknown: $\mathcal{L}_i$. Consequently, we have a total of
$2+(N-1)\times 1 = N+1$ unknowns.

Since we must have more equations than unknowns, we easily establish the
necessary condition:
\begin{equation}
(M-1)(N-1) \geqslant N+1 \quad\text{i.e.}\quad M(N-1) \geqslant 2N.
\end{equation}
\noindent
For $N$ = 2, 3, or 4, we easily find out that $M \geqslant$ 4, 3 or 
$\frac{8}{3}$, i.e. a minimum of 4, 3 or 3 filters is requested, respectively.

\medskip
\noindent
Therefore, observational data of multiply imaged quasars acquired through
several photometric filters are mandatory.

Note that if data obtained in $M$ filters come from several different 
observing runs, it is then safer to only use those colours formed between 
filters from the same run and hence we finally get less than $M-1$ useful 
colours. The above necessary condition may consequently not be any longer 
fulfilled.
\subsection{Colour-colour diagrams}
\label{colour-colour_diagrams}
Before applying the proposed method, we must make sure that in a colour-colour
diagram, and in accordance with Eq. (\ref{rel_colour}), the measurements of
the various lensed components are approximately located along a straight
line. Otherwise, it may mean that one or more of the conditions listed in
section \ref{generalities} are not fulfilled.

\begin{figure}[h]
\resizebox{\hsize}{!}{\rotatebox{270}{\includegraphics{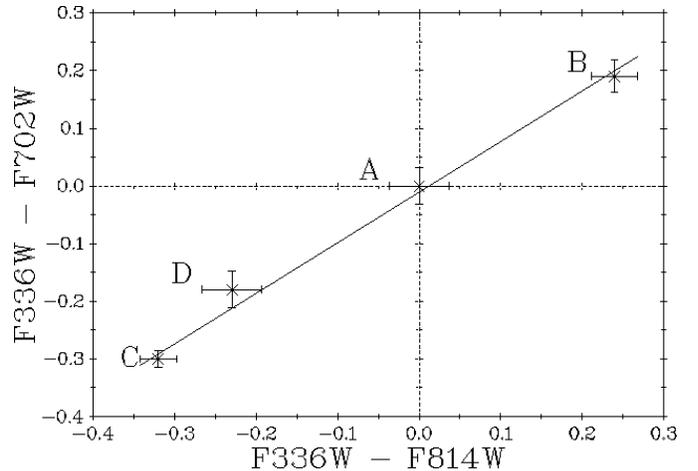}}}
\caption{Example of a colour-colour diagram for the case of 
\object{H1413+117}, constructed from observational data listed in Table 
\ref{phot_H1413}. Image A has been selected as the reference component.}
\label{diagram}
\end{figure}

Secondly, the differential reddening for most of the components has to be
significantly larger than the observational errors. Otherwise, the slope of the
straight line in the colour-colour diagram remains too uncertain. For example,
the components of the gravitational lens systems \object{PG~1115+080} (Kristian
\textit{et al.} \cite{Kristian}) and \object{B~1422+231} (Yee \& Ellingson \cite{Yee_Ellingson}; Yee \& Bechtold
\cite{Yee_Bechtold}) have very similar colours and hence these cases are not appropriate for
application of this method. Promising candidates are \object{H1413+117} (see an example
of colour-colour diagram in Fig. \ref{diagram}) and especially
\object{MG~0414+0534} which is known to be very stable (Moore \& Hewitt \cite{Moore}) and
probably not affected by significant microlensing effects (McLeod \textit{et
al.} \cite{McLeod}).

\subsection{Fitting of the measured colours}
The left side of equation (\ref{rel_colour}), which corresponds to an 
experimental relative colour, is directly derived from photometric
observations of the macro lensed quasar images.

The right side of equation (\ref{rel_colour}), and more explicitly the
quantity ``$\text{Ext}(\lambda_1',R_V) - \text{Ext}(\lambda_2',R_V)$", is
systematically computed for different contiguous values of $R_V$ and $z_l$
using the Galaxy extinction law from Cardelli \textit{et al.} 
(\cite{Cardelli}). The relative path lengths $\mathcal{L}_i$ are fitted using 
a routine that minimizes the following $\chi^2$:

\begin{equation}
\chi^2 = \sum \frac{(\text{Experimental colour} - \text{Fitted colour})^2}{\sigma^2_{\text{Experimental colour}}}.
\end{equation}

Figure \ref{carte_simul} presents such a $\chi^2$ map obtained from a set of simulated
data (see below).
\begin{figure}
\resizebox{\hsize}{!}{\rotatebox{270}{\includegraphics{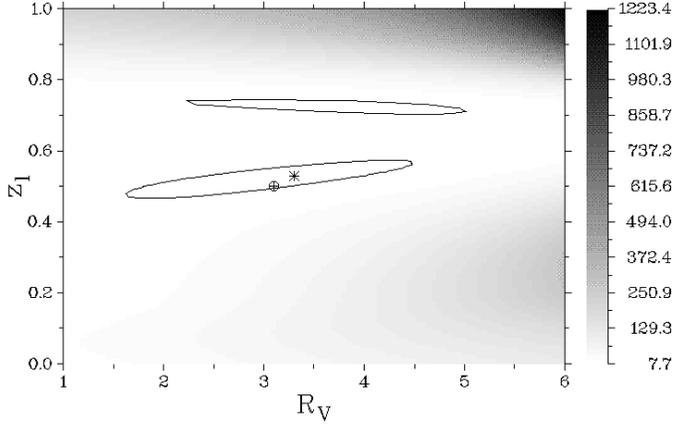}}}
\caption{$\chi^2$ map from simulations of '5 filters' observations, adopting 
$R_V=3.1$, $z_l=0.5$, $\mathcal{L}=0,0.6,1.3,1.1$ for components A, B, C and
D, respectively and a noise of 0.1 mag. for the simulated U, B, V, R and I 
magnitudes. The symbols $\oplus$ and $\ast$ correspond to the input and 
calculated values of the $z_l$ and $R_V$ parameters.}
\label{carte_simul}
\end{figure}
\section{Simulations}
In order to study the errors affecting the estimate of the lens redshift
using the fitting method described above, we have carried out a wide range of
Monte Carlo simulations. When doing these, we first fix the values of the
following parameters: the lens redshift $z_l$ within a defined interval, the 
parameter $R_V$ usually set to 3.1 - corresponding to the average galactic 
value -, the relative macro-amplification 
$\mathcal{A}_i = \frac{A_i}{A_{\text{ref}}}$ and the relative path length 
$\mathcal{L}_i$ through the 
deflector for each lensed component so that we can calculate precise relative 
magnitudes for as many different filters as we wish.
Then we add to these magnitudes a Gaussian noise of a certain level and apply
the above fitting method. We repeat these two last steps 100 times and
calculate the dispersion $\sigma$ in the estimate of the lens redshift
and the $R_V$ parameter.

\begin{figure}
\resizebox{\hsize}{!}{\includegraphics{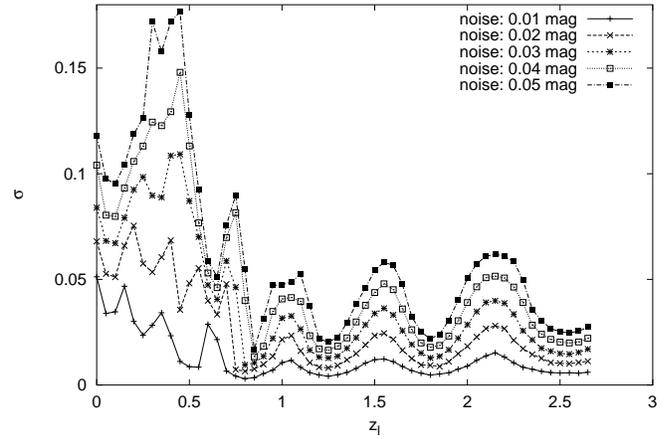}}
\caption{Lens redshift dispersion $\sigma$ versus the lens redshift $z_l$ for
five different levels of noise and adopting $R_V=3.1$, 
$\mathcal{L}=0,0.6,1.3,1.1$ for components A, B, C and D, respectively. The 
photometric observations have been simulated for the five U, B, V, R and I 
Johnson filters.}
\label{noise}
\end{figure}

\begin{figure}
\resizebox{\hsize}{!}{\rotatebox{270}{\includegraphics{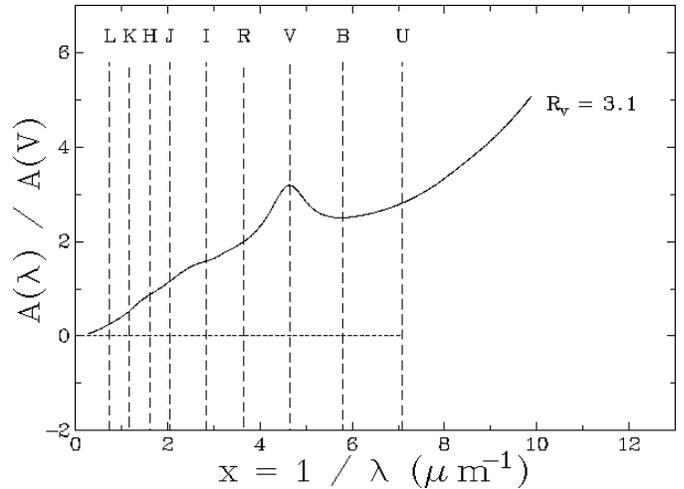}}}
\caption{Extinction law of Cardelli \textit{et al.} for $R_V = 3.1$. The 
redshifted ($z=1.55$) positions of selected filters are indicated.}
\label{plot_abs}
\end{figure}

Carrying out such simulations for different contiguous values of the
theoretical lens redshift and for levels of noise comprised between 0.01 mag.
and 0.05 mag., we obtain the curves illustrated in Fig. \ref{noise} for
which we used the set of the five U, B, V, R and I Johnson filters. We see
that the dispersion $\sigma$ affecting the estimate of $z_l$ is about 0.1
and even less at higher redshifts. The visible ``oscillations" in Fig. 
\ref{noise} correspond to the positions of the wavelengths of the different 
selected filters with respect to the redshifted 2175\,\AA\ absorption bump 
(see Cardelli \textit{et al.} \cite{Cardelli}). As an example, Fig. 
\ref{plot_abs} represents the extinction law of Cardelli \textit{et al.} 
(\cite{Cardelli}) as a function of the reciprocal wavelength 
$\frac{1}{\lambda}$ for $R_V=3.1$. The redshifted ($z=1.55$) positions of the 
U, B, V, R, I, J, H, K and L filters are indicated.

\begin{figure}
\resizebox{\hsize}{!}{\includegraphics{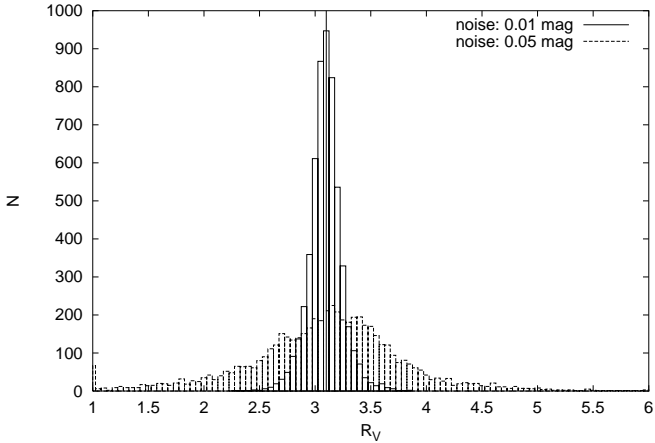}}
\caption{Histograms of the repartition of $R_V$ for two different levels of
noise and adopting $R_V=3.1$, $z_l$ from 0 to 2.65 and  
$\mathcal{L}=0,0.6,1.3,1.1$ for components A, B, C and D, respectively (see 
Fig. \ref{noise}).}
\label{histos_Rv}
\end{figure}

Furthermore, we have illustrated in Fig. \ref{histos_Rv} dispersions affecting
the derived values of $R_V$ for the cases of 5 filters and two different
levels of noise: 0.01 mag. and 0.05 mag. We see that, in this latter case, the
dispersion gets rather large.

\begin{figure}
\resizebox{\hsize}{!}{\includegraphics{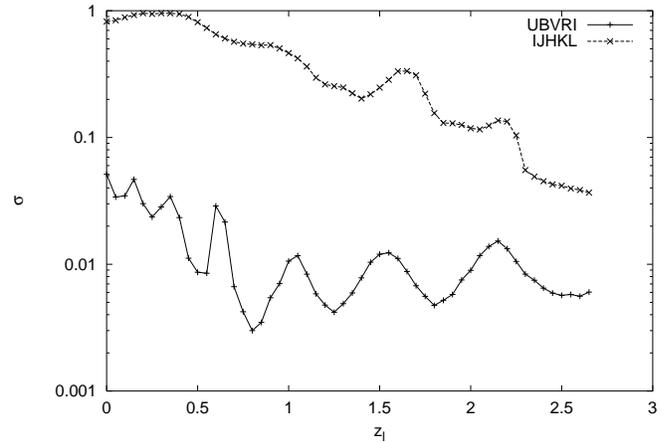}}
\caption{Lens redshift dispersion $\sigma$ versus the lens redshift $z_l$ for
two different sets of filters and adopting $R_V=3.1$, 
$\mathcal{L}=0,0.6,1.3,1.1$ for components A, B, C and D, respectively. A 
noise of 0.01 mag. has been used for the simulated photometric observations.}
\label{filters}
\end{figure}

Next, Fig. \ref{filters} shows on a logarithmic scale differences in the
lens redshift dispersion when using the U, B, V, R and I filters or the
infrared I, J, H, K and L filters. Since the method is sensitive to the
extinction gradient between two filter wavelengths, the much higher dispersion
derived for the infrared filters is mainly accounted for by the shallower 
shape of the
extinction law in the infrared range. Moreover, the lower extinction in the 
infrared range also affects the dispersion in the estimate of the lens 
redshift because the experimental relative colours are smaller and hence the 
effects of noise get more pronounced.

\begin{figure}
\resizebox{\hsize}{!}{\rotatebox{270}{\includegraphics{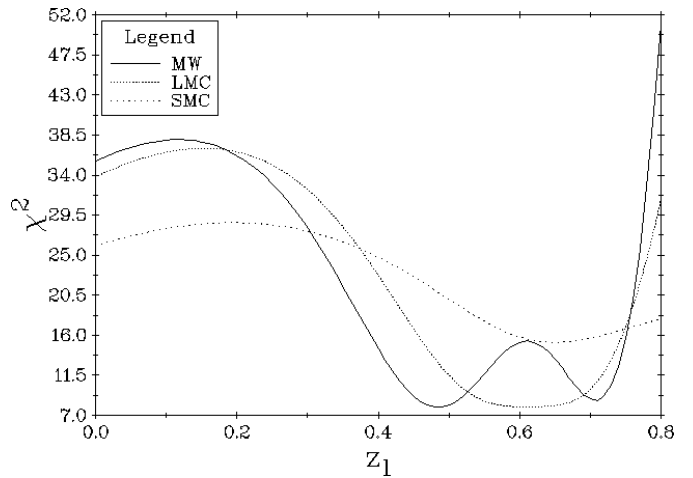}}}
\caption{$\chi^2$ curves from simulations of '5 filters' observations adopting 
three different extinction laws and $R_V = 3.1$, $z_l = 0.5$, 
$\mathcal{L}=0,0.6,1.3,1.1$ for components A, B, C and D, respectively and a 
noise of 0.1 mag. for the simulated magnitudes (see text and Fig. 
\ref{carte_simul}).}
\label{courbes_Pei}
\end{figure}

Finally, we have also used other extinction laws such those proposed by Pei 
(Pei \cite{Pei}) for the  Milky Way (``MW''), the Large Magellanic Cloud 
(``LMC'') and the Small Magellanic Cloud (``SMC''). From the same set of 
simulated data as in Fig. \ref{carte_simul}, Fig. \ref{courbes_Pei} shows the 
resulting $\chi^2$ curves. Comparable minima and lens redshift solutions are 
usually found for the MW and LMC laws. However, high quality photometric 
observations are mandatory in order to efficiently select the most likely lens 
redshift among acceptable solutions for the different extinction laws. Note 
that because of the noise in the simulated photometric data, the minimum for 
the MW curve in Fig. \ref{courbes_Pei} is found at $z_l = 0.48$, instead of 
the theoretical value $z_l = 0.5$.

\section{Applications to some real gravitational lenses}
\subsection{\object{2237+0305}}
\begin{figure}
\resizebox{\hsize}{!}{\rotatebox{270}{\includegraphics{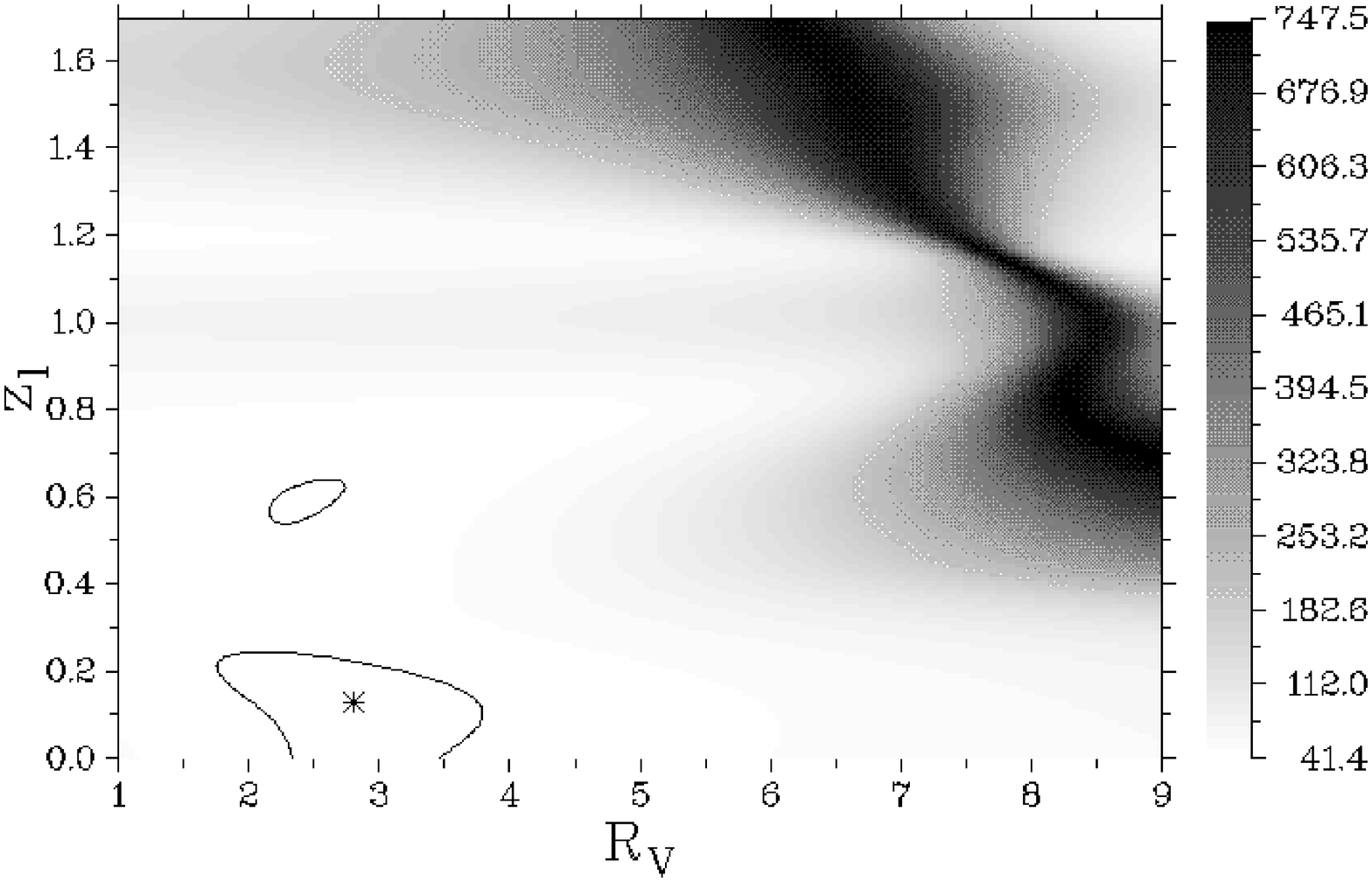}}}
\caption{$\chi^2$ map for \object{2237+0305} derived from the published sets of
photometric observations listed in Table \ref{phot_2237}. The $\chi^2$ minimum
corresponds to $z_l=0.13$ and $R_V=2.8$.}
\label{carte_2237}
\end{figure}

Figure \ref{carte_2237} represents the $\chi^2$ map for the \object{Einstein Cross} derived
from the published sets of photometric observations listed in Table \ref{phot_2237}. The
minimum is found for $z_l \simeq 0.13$. However, the redshift of this
gravitational lens is known to be 0.04 (Huchra \textit{et al.} \cite{Huchra}). The small discrepancy between these two
redshift values is very likely due to the fact that the data were obtained 
during two different runs when photometric quality was not good enough
($\sigma \geqslant 0.02$ mag. for approximately half the reported values in 
Table \ref{phot_2237}).
Moreover, this gravitational lens system is very likely affected by
microlensing effects on a time scale of several years (Irwin \textit{et al.} \cite{Irwin};
Corrigan \textit{et al.} \cite{Corrigan}; {\O}stensen \textit{et al.} \cite{Ostensen:2237}). The contours in Fig.
\ref{carte_2237} correspond to $\Delta\chi^2=1$ and delineate, at a confidence level
of $68 \%$, acceptable ranges for the values of $z_l$ and $R_V$. Very high
signal-to-noise photometric observations of the Einstein Cross, using several
filters during a single run and reliable tools to derive the photometry of 
the multiple QSO images, are badly needed to confirm that the different
experimental relative colours reported between the lensed components are
effectively, and solely, due to extinction effects in the deflector.
\begin{table*}
\caption{Photometry of \object{2237+0305}}
\begin{tabular}{ccccccc}
\hline
& & \multicolumn{4}{c}{Components} & \\
& & \multicolumn{4}{c}{\hrulefill} & \\
Date & Filter & A & B & C & D & Reference$^{\text{a}}$ \\ \hline
25/09/87 & g & 17.77$\pm$0.03 & 17.98$\pm$0.03 & 18.46$\pm$0.05 & 18.69$\pm$0.06 & 1 \\
25/09/87 & r & 17.62$\pm$0.02 & 17.77$\pm$0.03 & 18.06$\pm$0.04 & 18.40$\pm$0.06 & 1 \\
25/09/87 & i & 17.25$\pm$0.02 & 17.38$\pm$0.02 & 17.61$\pm$0.03 & 17.98$\pm$0.04 & 1 \\
10-11/10/95 & B & 17.626$\pm$0.009 & 17.741$\pm$0.008 & 18.881$\pm$0.020 & 19.111$\pm$0.033 & 2 \\
10-11/10/95 & V & 17.278$\pm$0.004 & 17.428$\pm$0.009 & 18.389$\pm$0.009 & 18.734$\pm$0.022 & 2 \\
10-11/10/95 & R & 17.093$\pm$0.004 & 17.274$\pm$0.004 & 18.109$\pm$0.008 & 18.441$\pm$0.010 & 2 \\
10-11/10/95 & I & 16.966$\pm$0.006 & 17.192$\pm$0.007 & 17.910$\pm$0.015 & 18.206$\pm$0.020 & 2 \\
\hline
\end{tabular}
\\
$^{\text{a}}$(1) Yee \cite{Yee}; (2) Burud et al. \cite{Burud} (CLEAN algorithm).
\label{phot_2237}
\end{table*}
\subsection{\object{H1413+117}}
Figure \ref{carte_H1413} shows the $\chi^2$ map for the Cloverleaf. For this 
case, we used photometric data obtained through 7 filters but from two distinct
observing runs (see Table \ref{phot_H1413}). We see that a minimum is found for $z_l
\simeq 1.15$ and $R_V \simeq 9$. This latter value looks quite unexpected and
we also conclude here that very high signal-to-noise photometric data of the
Cloverleaf, using several filters during a single run, are badly needed to
confirm that the reddening observed between the macro-lensed QSO images is
essentially due to extinction effects. For this gravitational lens system,
numerous lens redshifts from identified absorption lines have been proposed:
0.31 and 0.75 (Afanas'ev \textit{et al.} \cite{Afanasev}), 1.44 (Magain \textit{et al.}
\cite{Magain}), 1.66, 1.87 and 2.07 (Hazard \textit{et al.} \cite{Hazard}; Drew \& Boksenberg
\cite{Drew}; Turnshek \textit{et al.} \cite{Turnshek88}) but the real lens redshift still remains
unknown.
\begin{table*}
\caption{Photometry of \object{H1413+117}}
\begin{tabular}{ccccccc}
\hline
& & \multicolumn{4}{c}{Components} & \\
& & \multicolumn{4}{c}{\hrulefill} & \\
Date & Filter & A & B & C & D & Reference$^{\text{a}}$ \\ \hline
02/08/94 & B & -1.06$\pm$0.02 & -0.82$\pm$0.02 & -0.88$\pm$0.02 & -0.70$\pm$0.02 & 1 \\
02/08/94 & V & -0.41$\pm$0.01 & -0.23$\pm$0.01 & -0.20$\pm$0.01 & -0.06$\pm$0.01 & 1 \\
02/08/94 & R & -0.08$\pm$0.01 &  0.10$\pm$0.01 &  0.18$\pm$0.01 &  0.28$\pm$0.01 & 1 \\
02/08/94 & I &  0.19$\pm$0.01 &  0.36$\pm$0.01 &  0.48$\pm$0.01 &  0.62$\pm$0.01 & 1 \\
22/12/94 & F336W & 19.03$\pm$0.03 & 19.37$\pm$0.02 & 19.01$\pm$0.01 & 19.20$\pm$0.03 & 2 \\
22/12/94 & F702W & 18.40$\pm$0.01 & 18.55$\pm$0.02 & 18.68$\pm$0.01 & 18.75$\pm$0.01 & 2 \\
22/12/94 & F814W & 18.66$\pm$0.02 & 18.76$\pm$0.02 & 18.96$\pm$0.02 & 19.06$\pm$0.02 & 2 \\
\hline
\end{tabular}
\\
$^{\text{a}}$(1) {\O}stensen et al. \cite{Ostensen:H1413}; (2) Turnshek et al. \cite{Turnshek97}.
\label{phot_H1413}
\end{table*}
\begin{figure}
\resizebox{\hsize}{!}{\rotatebox{270}{\includegraphics{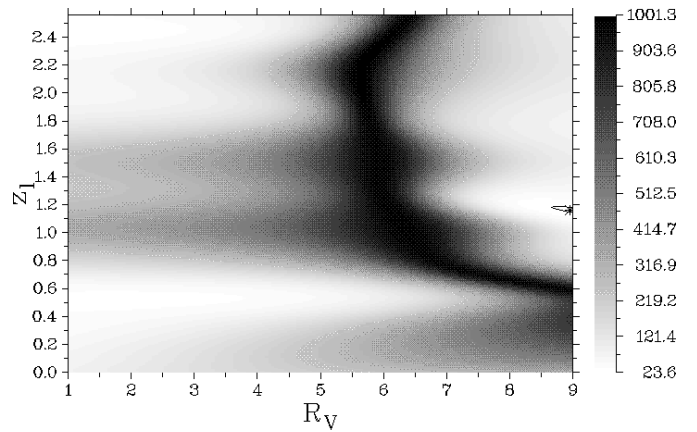}}}
\caption{$\chi^2$ map for \object{H1413+117} derived from the published sets of
photometric observations listed in Table \ref{phot_H1413}. The $\chi^2$ minimum
corresponds to $z_l=1.15$ and $R_V \simeq 9$.}
\label{carte_H1413}
\end{figure}
\subsection{\object{MG 0414+0534}}
\begin{figure}
\resizebox{\hsize}{!}{\rotatebox{270}{\includegraphics{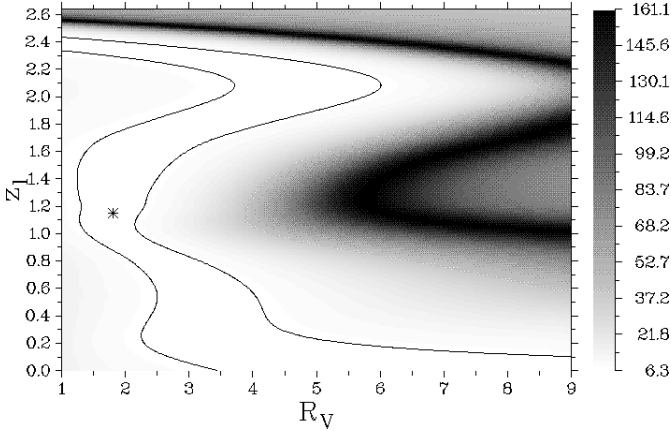}}}
\caption{$\chi^2$ map for \object{MG 0414+0534} derived from the published sets of
photometric observations listed in Table \ref{phot_MG0414}. The $\chi^2$ minimum
corresponds to $z_l=1.15$ and $R_V=1.8$.}
\label{carte_MG0414}
\end{figure}

Figure \ref{carte_MG0414} shows the $\chi^2$ map for \object{MG 0414+0534} 
derived from the published sets of photometric observations listed in Table 
\ref{phot_MG0414}. Because we only have 2 colours, including the near infrared 
colour H$-$K (see Table \ref{phot_MG0414}), the estimate of the lens 
redshift by this method remains very uncertain (see the very widespread 
contour in Fig. \ref{carte_MG0414}). Due to the known flux stability of 
\object{MG~0414+0534} in the radio (Moore \& Hewitt \cite{Moore}), this 
gravitational lens system constitutes one of the most promising targets for an
estimate of the lens redshift from the observed reddening of its lensed
components. Microlensing effects must be insignificant for this object (McLeod
\textit{et al.} \cite{McLeod}). High quality and simultaneous photometric data 
from the visible up to the near infra-red are very much needed for this 
interesting system.
\begin{table*}
\caption{Photometry of \object{MG 0414+0534}}
\begin{tabular}{ccccccc}
\hline
& & \multicolumn{4}{c}{Components} & \\
& & \multicolumn{4}{c}{\hrulefill} & \\
Date & Filter & A & B & C & D & Reference$^{\text{a}}$ \\ \hline
10/03/93 & H & 15.31$\pm$0.05 & 15.51$\pm$0.05 & 16.28$\pm$0.01 & 17.16$\pm$0.02 & 1 \\
10/03/93 & K & 14.34$\pm$0.05 & 14.56$\pm$0.05 & 15.36$\pm$0.01 & 16.24$\pm$0.02 & 1 \\
08/11/94 & F675W & 22.760$\pm$0.008 & 23.756$\pm$0.016 & 23.488$\pm$0.012 & 24.258$\pm$0.022 & 2 \\
08/11/94 & F814W & 20.595$\pm$0.003 & 21.407$\pm$0.004 & 21.363$\pm$0.004 & 22.212$\pm$0.007 & 2 \\
\hline
\end{tabular}
\\
$^{\text{a}}$(1) McLeod et al. \cite{McLeod}; (2) Falco et al. \cite{Falco}.
\label{phot_MG0414}
\end{table*}
\section{Generalization}
From Eq. (\ref{flux}) and considering the $j$ component relative to the \emph{ref}(-erence) component, we may also write:
\begin{equation}
\frac{F_{\lambda}^j}{F_{\lambda}^{\text{ref}}} = \frac{A_j}{A_{\text{ref}}}\,e^{-\text{Ext}(\lambda',R_V)[L_j - L_{\text{ref}}]} =
\mathcal{A}_j\,e^{-\text{Ext}(\lambda',R_V)\mathcal{L}_j}
\end{equation}
where $\mathcal{A}_j = \frac{A_j}{A_{\text{ref}}}$.
From this equation, we straightforwardly derive:
\begin{equation}
m_{\lambda,j}-m_{\lambda,\text{ref}} =
\frac{2.5}{\ln(10)} \, \mathcal{L}_j \, [\text{Ext}(\lambda',R_V)] \,
-2.5 \log(\mathcal{A}_j).
\label{rel_mag}
\end{equation}
We see that in fact, one can also directly work with the relative magnitudes
$m_{\lambda,j}-m_{\lambda,\text{ref}}$ but that the relative macro lens amplifications
$\mathcal{A}_j$ then appear as additional unknowns which also need to be
fitted. Following a reasoning similar to that adopted in section
\ref{constraints}, the necessary condition is also found to be:
$M(N-1) \geqslant 2N$, where $M$ and $N$ have the same meaning. Before 
applying this generalized method to real photometric observations, one should 
make sure that intrinsic or extrinsic (cf. induced by micro-lensing) light 
variations are not affecting the individual lensed components. These 
conditions are of course more difficult to be checked than those imposed on 
the observed colours. 

Finally, it is also straightforward to establish from Eq. (\ref{rel_mag}) that 
the redshifted extinction law $\text{Ext}(\lambda')$ may be directly retrieved 
as a function of $\lambda'$ from the observed spectra of at least three 
multiple lensed QSO images. Very high signal-to-noise, low spectral
resolution, VLT observations of gravitational lens systems like \object{H1413+117} and
\object{MG~0414+0534} should enable one to derive such a redshifted extinction 
law and compare it with the known MW, LMC and SMC ones.
\section{Conclusions}
Monte Carlo simulations show that we need very accurate photometric data
(typically $\sigma \simeq 0.01$ mag.), in several ($\geqslant 4 - 5$) filters, 
in order to estimate the lens redshift from the differential reddening 
observed between the lensed 
images of a multiply imaged quasar. Furthermore, the macro-lensed images must 
not be affected by light contamination from the lensing galaxy, by microlensing
effects or by intrinsic colour variations reflected with a time dependence
between the lensed components. Unfortunately, all these conditions are rarely
fulfilled at the same time and existing photometric data are often not good
enough or not sufficiently reliable. Consequently, accurate photometric 
observations of gravitational lenses obtained in several filters during the 
same run are badly needed.
Finally, we have seen that the redshifted extinction law of the lensing galaxy
could be directly retrieved by means of a similar method provided that we may 
record the individual spectra of at least 3 lensed components, over a wide 
wavelength range, assuming that no light contamination, no microlensing 
effects, \ldots affect the individual lensed components. Very good such 
candidates to be observed with the VLT are \object{H1413+117} and \object{MG~0414+0534}.

\begin{acknowledgements}
CJ is supported by contract ARC94/99-178 ``Action
de Recherche Concert\'ee de la Communaut\'e Fran\c{c}aise (Belgium)" and
``P\^ole d'Attraction Interuniversitaire, P4/05 (SSTC, Belgium)". JS would
like to acknowledge support from the SSTC/PRODEX project ``Observations of
gravitational lenses''.
\end{acknowledgements}

\noindent
\textit{Note added in proof:} On the basis of spectroscopic observations 
obtained for \object{MG~0414+0534} with the KECK telescope (see preprint 
astro-ph/9809063 submitted on 5 September 1998), J.L. Tonry and C.S. Kochanek 
report the redshift determination ($z_l\simeq 0.96$) of the very faint 
gravitational lens. Based upon preliminary photometric observations of this 
system, our own redshift estimate is found to be in reasonable agreement with 
this value (see Fig. \ref{carte_MG0414}).

\end{document}